\documentstyle[11pt,epsfig,epsf]{article}
\newcommand\C{\v Cerenkov }
\def\jnfont{\rm}
\def\NPB#1,{{\jnfont Nucl.\ Phys.\ }{\bf B#1},}
\def\PLB#1,{{\jnfont Phys.\ Lett.\ B }{\bf #1},}
\def\PRD#1,{{\jnfont Phys.\ Rev.\ D }{\bf #1},}
\def\PRL#1,{{\jnfont Phys.\ Rev.\ Lett.\ }{\bf #1},}
\def\ZPC#1,{{\jnfont Z.~Phys.\ C }{\bf #1},}
\def\ETslash{\not{\hbox{\kern-4pt $E_T$}}}
\let\to=\rightarrow
\begin{document}

\thispagestyle{empty}

\begin{flushright}
VLBL Study Group-H2B-2\\
IHEP-EP-2001-02\\
Ames-HET-01-08\\
%E$_\nu = 20$ GeV \\
%new input parameters\\
\today 
\end{flushright}
\vskip 3ex
\begin{center}
\vskip 2ex
\LARGE
On the Optimum Long Baseline for the Next Generation \\
Neutrino Oscillation Experiments  \\
\normalsize
\vskip 4ex
Yifang Wang
\vskip 1ex
Institute of High Energy Physics, 
Chinese Academy of Sciences \\
\vskip 1ex
Beijing 100039, China\\
\vskip 4ex
Kerry Whisnant and Bing-Lin Young\\
\vskip 1ex
Department of Physics and Astronomy, 
Iowa State University \\
\vskip 1ex
Ames, Iowa 50011, U.S.A.\\
\end{center}

\abstract{  }

For high energy long baseline neutrino oscillation experiments, we 
propose a Figure of Merit criterion to compare the statistical quality 
of experiments at various baseline lengths under the condition of
identical detectors and a given neutrino beam.  We take into account 
all possible experimental errors under general consideration.  In this
way the Figure of Merit is closely related to the usual statistical criterion 
of number of sigmas.  We use a realistic neutrino beam for an entry 
level neutrino factory and a possible superbeam from a meson source 
and a 100 kt detector for the calculation.  We considered in detail
four baseline lengths, 300 km, 700 km, 2100 km and 3000 km, 
in the neutrino energy range of 0.5-20 GeV for a 20 GeV entry 
level neutrino factory and a 50 GeV superbeam.  We found that the 
very long baselines of 2100 km and 3000 km are preferred for the
neutrino factory according to the figure of merit criterion.  Our 
results also show that, for a neutrino factory, lower primary muon 
energies such as 20 GeV are preferred rather than higher ones such as 30 
or 50 GeV.   For the superbeam, the combination of a long baseline
such as 300 km and a very long baseline like 2100 km will form a 
complete measurement of the oscillation parameters besides the
CP phase.  To measure the CP phase in a superbeam, a larger
detector (a factor 3 beyond what is considered in this article) and/or
a higher intensity beam will be needed to put some significant 
constraints on the size of the CP angle.

\newpage
\section{Introduction}

\vskip 2ex

Neutrino oscillations are to date the only experimental indication of 
physics beyond the standard model that may further expand the 
horizon of our most basic knowledge of nature. 
%Historically neutrinos have played mostly passive roles in particle
%physics research.  However, the observation of oscillations in a host
%of solar and atmospheric neutrino experiments, such as the 
%Super-Kamiokande experiment~\cite{superK} and the various other 
%corroborating experiments~\cite{experiments},
%have changed
%significantly the role of the neutrino.  This has profound implications 
%for particle physics,
%astrophysics and cosmology.  Neutrinos are the end product of the
%decays of almost all elementary particles, observed or proposed.
%They fill the universe as a relic background radiation similar to
%the photon as a fingerprint of the early history of the universe.  Now
%they are taking the center stage in particle physics research and
%their heightened fascination by physicists has just began. 
%
Although the existing data from
Super-Kamiokande experiment~\cite{superK} and the various other 
corroborating experiments~\cite{experiments}
offer very strong indications of neutrino
oscillations, the oscillation parameters have not been determined with 
sufficient accuracy and an appearance experiment have not been 
convincingly performed.  
%The unique signature of the flavor transmutation, i.e., the 
%appearance of a flavor different from the original one, has been
%observed by one experiment but not confirmed yet by others.  
In the ongoing neutrino oscillation experiments and those of the
next generation under construction, some of the parameters will be
probed with greater
accuracy  but others may not yet be accessible.  Hence more experiments 
designed to look for the missing information, to carry out appearance 
experiments, and to probe the known parameters with even greater accuracy
are desirable.    

%The neutrino oscillation is a system with a limited number of degrees
%of freedom, yet it exhibits a multitude of interesting phenomena. In 
%the 3-flavor scenario, the system consists of 2 mass square differences 
%(MSD), three mixing angles and one measurable CP phase that determine 
%the different survival and appearance probabilities~\cite{BDWY, lecture}.  
%The numerous neutrino experiments, solar,
%atmospheric, reactor, and short baseline, are mostly looking at the
%survival probabilities and often use neutrino beams that exist in nature.
%In most cases there is no way to tune the neutrino beams for more desirable
%experimental results.  Hence it is difficult to obtain all the oscillation 
%parameters for the entire mixing matrix, either because the statistics
%are low, or the energy and distance are not suitable.  In long baseline 
%(LBL) experiments,
%the neutrino beams are produced in an accelerator according to definite  
%physics criteria so that the experiment can be conducted in a controlled
%fashion.  Ideally, the distance between the neutrino source and the
%detector can be chosen to maximize the physics output.  The distance can
%be hundreds or thousand kilometers to allow high energy neutrino beams
%to be used and still offer a suitable $L/E_\nu$ ratio.  The LBL experiments
%promise to allow for a detailed analysis of the oscillation parameters so as
%to provide a complete picture of the neutrino oscillations.
%It is therefore very important to investigate the optimum baseline 
%of the LBL experiment for the future choice of baseline length
%and neutrino energy.

In this article we propose a criterion for the determination in general terms 
and on a global scale the overall quality of long baseline (LBL) experiments.
We focus on high energy neutrino beams in the range of 0.5- 20 GeV 
with the possibility of $\tau$ production and CP violation measurement, 
and distances from 300 km to 3000 km.   At the lower end, actually 295 km 
baseline, is the proposed oscillation experiment~\cite{J2K} using the
neutrino beam from the newly approved high intensity 50 GeV proton synchrotron 
in Japan called HIPA~\cite{HIPA} with the Super-Kamiokande detector
or its updated version.  The 700 km baseline is close to the first group of
next generation LBL experiments: MINOS~\cite{MINOS}, 
ICARUS~\cite{ICARUS} and OPERA~\cite{OPERA}, all having  a  
baseline of about 730 km.  The 2100 km distance is a possibility for
a very long baseline LBL experiment, called H2B, 
under discussion~\cite{H2B,Japanesegroup}.  The neutrino beam would be 
from HIPA and 
the detector, tentatively called the Beijing Astrophysics and Neutrino Detector
(BAND), will be located in Beijing, China.  The 3000 km baseline has been
discussed extensively in neutrino factory studies~\cite{Albright} although
particular accelerator and detector sites have not been formally identified.

To compare different experiments is usually difficult. It requires knowledge 
of the actual detectors, their different systematics, and their neutrino beams.
This is beyond the scope of the present work.  So for a possible yet
meaningful comparison we assume that all experiments have identical 
detectors and share a given neutrino beam. We believe that our present 
idealized approach allows us to make a meaningful comparison of  the 
capabilities of the various baseline experiments in general terms and will 
be useful for the choice of the most suitable baseline length with a 
given neutrino beam.

In Sec. 2 we present some of the fundamentals of LBL experiments which
include the two types of neutrino beams and  their beam fluxes that we 
will use in our calculations. We also
list the neutrino charge current cross sections. In Sec. 3 we propose the
figure of merit as a criterion to determine the quality of an experiment.
Section 4 discusses the figure of merit of the various measurements in the 
three-neutrino scheme at four baselines: 300 km, 700 km, 
2100 km and 3000 km.   In practice, for the appearance measurement a
superbeam will measure $\nu_\mu\rightarrow \nu_e$, $\nu_\tau$ or 
$\bar{\nu}_\mu\rightarrow \bar{\nu}_e$, $\bar{\nu}_\tau$ 
while a neutrino factory can also measure $\nu_e\rightarrow \nu_\mu$,
$\nu_\tau$ or 
$\bar{\nu}_e\rightarrow \bar{\nu}_\mu$, $\bar{\nu}_\tau$. 
In order to compare the neutrino factory and the neutrino superbeam, 
in this paper we only use the $\nu_\mu$ and $\bar{\nu}_\mu$ 
beams in our consideration.  The measurements considered include 
$\nu_e$ and $\nu_\tau$ appearance, the matter effect, leading oscillation
parameters, sign of the leading neutrino mass-square difference, and 
the CP violation effect.  Section 5 contains a brief summary and some 
pertinent comments, in particular, the comparison of neutrino factories 
of different primary muon energies.

\section{Fundamentals of LBL experiments}

%Long baseline neutrino oscillation experiments are not conventional high
%energy physics experiments. 
Due to the extremely weak interaction cross
section of neutrinos and beam divergence at a long distance from the source,
both the neutrino beam intensity and the detector mass
must be maximized in order to have the desired statistics. New technologies
for both accelerators and detectors may be required. In the following we
discuss briefly some of the fundamentals of the LBL experiment which 
will be used in our calculations.

\noindent
\subsection{ Neutrino beams}

%There are two kinds of accelerator neutrino beams: the neutrino
%factory from muon decays and the conventional neutrino beam from meson
%decays. While meson-neutrino beams have been built by many laboratories
%and the remaining technological
%challenge is to increase the total power of the primary proton beam, the 
%neutrino factory is a completely new concept and there are still 
%a host of technical issues to be worked out.
\vskip 2ex

{\bf Neutrino factory}:  The neutrino factory delivers a neutrino beam
which contains comparable amount of $\nu_\mu$ ($\bar{\nu}_\mu$) and
$\bar{\nu}_e$ ($\nu_e$) obtained from the $\mu^-(\mu^+)$ decay in a 
$\mu$-storage ring. 
\begin{equation}
\mu^-(\mu^+)\to \nu_\mu(\bar{\nu}_\mu) + \bar{\nu}_e(\nu_e)+ e^-(e^+) 
\end{equation}
%Note that 
The presence of both muon and electron neutrinos is not a
problem because they have opposite sign charged leptons in charge 
current reactions and 
therefore will not be a background to each other if the detector can 
distinguish between positive and negative electric charges.  A description 
of the neutrino factory can be found in \cite{Geer}.

The neutrino flux at baseline L from a neutrino factory of unpolarized
muon of energy $E_\mu$ is given by
\begin{equation}
{d\Phi\over dE_\nu}= \left\{ \begin{array}{lll}
    2x^2_f (3-2x_f){n_0 \gamma^2\over \pi L^2 E_\mu} & for & \nu_\mu \\
    12x^2_f(1-x_f){n_0 \gamma^2\over \pi L^2 E_\mu} & for & \nu_e 
    \end{array} \right.
\label{eq:nuf}
\end{equation}
where $x_f = E_\nu /E_\mu$, $n_0$ is the number of useful decaying muons,
and $\gamma = E_\mu/m_\mu$ with $E_\mu$ and $m_\mu$ being respectively
the energy and mass of the muon.  
%The average neutrino energies are given by 
%\begin{eqnarray}
%\langle E_{\nu_\mu}\rangle &=& 0.7E_\mu \nonumber \\
%\langle E_{\nu_e}\rangle &=& 0.6E_\mu
%\end{eqnarray}
Two scenarios of the number of neutrinos in a beam have been considered:
$n_0=6\times 10^{19}/$year for an entry level factory and 
$n_0=6\times 10^{20}/$year for a high performance factory.   
For a discussion of the neutrino beam spread in a 
neutrino factor together with some sample plots, see Ref. \cite{BGW}. 
 
\vskip 2ex
{\bf Meson-neutrino superbeam}: Neutrinos from a meson source are 
obtained from decays of pions and kaons produced by collisions of the 
primary proton with the nuclear target.  
The secondary meson beam produced by the collision is sign selected
and then focused by a magnetic field. The meson beam is then transported
to a vacuum decay pipe, whose length depends on the desired energy of
the neutrino.% and finally striking a hadron absorber further downstream.   
The primary neutrino or anti-neutrino beam consists mostly of the muon
flavor from $\pi^\pm$ and $K^\pm$ decays in the decay pipe but some
impurities of electron neutrinos are expected due to a finite branching
ratio of $\pi^\pm$, $K^\pm$ and $\mu^\pm$ decaying into electron
neutrinos. For example, the NuMI muon neutrino beam at Fermilab
contains 0.6\% electron neutrinos.   The neutrino beam energy profile          %%%
is more complicated.  In general, it can be a wide-band beam covering a        %%%
broad range of energies, or narrow band beam with a selected, 
well-defined narrow range of energy.   

The neutrino flux at the detector site is determined by the baseline L, the
number of primary protons on target (POT), the proton energy $E_p$, and 
the neutrino energy $E_\nu$. The following empirical formula~\cite{malensek} 
describes the meson production from a proton beam on a nuclear target:
%
%  YFW:
%
% A better formula will be available from A. Marchioni, this is just 
%temporary.
\begin{eqnarray}
x_M{d\sigma\over dx_M} & = & 2\pi\int x_ME_p {d^3\sigma\over dp^3}
                  P_tdP_t    \nonumber\\
                   & = & 2\pi\int {B(1-x_M)^A{1+5e^{-Dx_M}\over 
                         (1+P^2_{t}/C)^4}}P_tdP_t \nonumber\\
                   & = & N_M (1-x_M)^A(1+5e^{-Dx_M})
\end{eqnarray}
where %$E_p$ is the proton beam energy, 
$N_M$ is a normalization factor, $p$ the proton 3-momentum, $x_M$ 
the Feynman x-variable defined as the momentum of the secondary
meson divided by the momentum of proton.  A, B, C, and D are numerical
parameters which are different for different secondary particles.  
Table~1.  %\ref{tab:mesonpara} 
gives their fitted values for $\pi^+, \pi^-, K^+$, and $K^-$, taken from
Ref.\cite{malensek}. 
By taking the general property of Eq. (2), i.e., the $E^2/L^2$ behavior,
to account for the transverse momentum spread due to pion decays,
the envelope of the neutrino flux which is the maximum flux at each
energy, can be written as,
% For a wide band beam, we can take Eq.(\ref{eq:nuf}), 
% i.e., the $\mathrm E_\nu^2/L^2$ behavior, to account for the  
% transverse momentum spread due to pion decays. 
% The envolope of the neutrino flux can then be written as                    %%%      
\begin{eqnarray}
\Phi (E_\nu,L) \propto {E_\nu(1-x_\nu)^A(1+5e^{-x_\nu D}) 
\over L^2}
\label{eq:beam}
\end{eqnarray}
where we take $x_M \simeq x_\nu \equiv 2E_\nu/E_p$. 
The wide or narrow band beam can then be selected from this          %%%
envelope of the total beam.   It is interesting to note 
that the simple formula given in Eq. ~(\ref{eq:beam})                        %%%   
can account for the beam design of MINOS at various 
energies as discussed in Ref.~\cite{Albright}.   We should remark that there is a 
more complete treatment~\cite{neumesonbeam} for the energy spectrum of the 
meson-neutrino beam.  However the difference with the above expression for 
$E_\nu > 1$ GeV is very small. Owing to its simpler form, we use 
the above expression in the following calculations.

\begin{table}[hbtp!]
\begin{center}
\begin{tabular}{|l|c|c|c|c|}
\hline
            &  A  & B & C & D \\
\hline
$\pi^+$     & 2.4769 & 5.6817E-2 & 0.57840 & 3.0894 \\
$\pi^-$     & 3.5648 & 5.0673E-2 & 0.68725 & 5.0359 \\
$K^+$       & 1.7573 & 6.3674E-3 & 0.81771 & 5.6915 \\
$K^-$       & 5.4924 & 4.1712E-3 & 0.89038 & 2.2524 \\
\hline
\end{tabular}
\parbox{5.75in}{\caption{Numerical parameters for meson-neutrino energy 
                           spectrum}}
\label{tab:mesonpara}
\end{center}
\end{table}
%%%%
            
As a function of $E_\nu$ and $L$ the flux of an entry level neutrino factory
is given by 
\begin{equation}
\Phi_f(E_\nu,L)=2 x_f^2(3-2x_f){n^{(f)}_0\gamma_\mu^2\over 
                             \pi L^2 E_\mu}
            \times 6\times 10^{22}
\label{eq:neutrinofactory}
\end{equation}
per thousand tons (kt) per year and per cm$^2$, where 
$n^{(f)}_0 = 6\times 10^{19}$/year .  We will take the primary muon energy to
be $E_\nu = 20$ GeV.
            
For the neutrino beam of the meson source we take the flux as
\begin{equation}
\Phi_m(E_\nu,L) = (1-x_\nu)^A (1+ 5e^{-D x_\nu}){E\over L^2} 
                   n^{(m)}_0\times 6\times 10^{22}.
\label{eq:mesonsource}
\end{equation}
%where the $\pi^+$ parameters are used. 
The normalization factor 
$n^{(m)}_0$ is determined by the total number of $\nu_\mu$'s in the HIPA 
broad band neutrino beam for $E_p = 50$ GeV at $L=295$ km \cite{J2K}.  
This gives $n^{(m)}_0 = 13.7\times 10^{19}$ in the proper units. 
We note that Eq. (6) gives only the envelope of the flux, not the actual
shape of the energy spectrum.  A more precise simulation can be made 
once the exact beam profile is given. 
%We note 
%that the shape of the energy spectra given in Eq.~(\ref{eq:mesonsource}) 
%may not match exactly that of the HIPA designed broad band beam.  The    
%form is intended for illustration, not as a precise simulation.
%More precise simulation can be made once the exact beam profile is given.  
\subsection{Interaction cross sections}

The detection of the neutrino flavor is through the charge current
interaction. For a neutrino energy which is small compared to the mass of
the W-boson, the charge current cross sections for the electron and muon
neutrino are given by
\begin{eqnarray}
\sigma^{(e,\mu)}_{\nu N} & = & 
         0.67\times 10^{-38}{\rm cm}^2 E_\nu(\rm GeV)  \\
\label{eq:crossp}
%\end{eqnarray}
%\begin{eqnarray}
\sigma^{(e,\mu)}_{\bar{\nu} N} & = & 
         0.34\times 10^{-38}{\rm cm}^2 E_\nu(\rm GeV)  
\label{eq:crossn}
\end{eqnarray}

For the tau neutrino, the above expression is subject to a threshold 
suppression.  The threshold for the production of the $\tau$ is
$E_\tau = 3.46$ GeV. 
%\begin{equation}
%E_T = m_\tau + {m_\tau^2\over 2m_N}= 3.46~{\rm GeV}
%\end{equation}
The charge current production cross section of the $\tau$ is usually given 
numerically as a function of the neutrino energy.  We fit the numerical 
cross sections \cite{taunumerical} from the $\tau$ threshold to 100 GeV 
and  obtained the following expression
\begin{equation}
\sigma^{(\tau)}_{\nu N}/\sigma^{(\mu)}_{\nu N} = {(E_\nu - E_T)^2\over 
                c_0 + c_1 E_\nu + c_2 E^2_\nu}\theta(E_\nu-E_T), 
\end{equation}
where $c_0=-84.988$, $c_1=18.317$, and $c_2=1.194$. As shown in 
Fig.~1,  %\ref{fig:0taucs}, 
the fit, which is valid for  $E_\nu \geq 4.0$ GeV, is 
good to within 3\%.  The difference occurs mostly in the neutrino energy 
region of 20-40 GeV.  

%%%%%%%%%%%%%%%%%%%%%%%%%%%%%%%%
\begin{figure}[htbp]
\begin{center}

\mbox{\epsfig{file=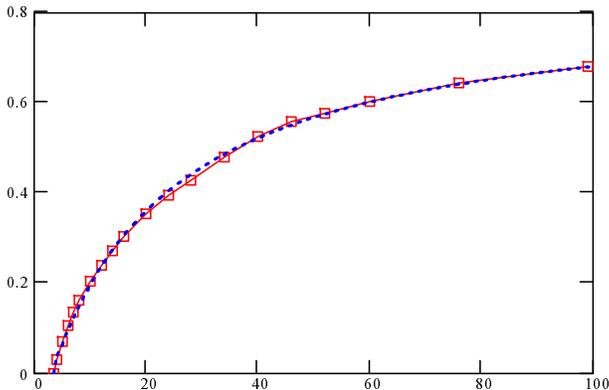, height=6cm, clip=}}
%\epsfxsize=10cm \epsfbox{taucs.eps}
%\addtolength{\captionwidth}{-2.0in}
\parbox{5.75in}{\caption[]{A fit of the numerical ratio of the $\tau$ to $\mu$ 
                           charge current production cross section from the $\tau$ 
                           threshold to 100 GeV.  The boxes are the data points and 
                           the dotted curve is the fit. The horizontal axis is the neutrino 
                           energy in GeV and the vertical axis is the ratio of $\tau$ to 
                           $\mu$ production cross sections. }}
\label{fig:0taucs}
\end{center}
\end{figure}

\section{Statistics and Figure of Merit}

%%%%(2)
Let us consider the oscillation: $\nu_\alpha\rightarrow \nu_\beta$.
We define two differential number of particle distributions as functions 
of the neutrino energy and baseline length.  In the following we will 
drop the subscript $\nu$ and use $E$ for the neutrino energy :
\begin{eqnarray}
N_{0\alpha}(E,L) &=& \Phi (E,L) K_t T_y \sigma_{\alpha}(E)
\\
N_{0\beta}(E,L)  &=&  \Phi (E,L) K_t T_y \sigma_{\beta}(E)
\end{eqnarray}
where $K_t$ is in kilotons and $T_y$ in years. $\Phi (E,L)$ is the flux 
of the $\nu_\alpha$ 
%beam at energy E and 
%distance L for either the neutrino factory or meson beam as 
and given in 
Eqs.~(\ref{eq:neutrinofactory}) and (\ref{eq:mesonsource}), 
$\sigma_{\alpha}(E)$ and $\sigma_{\beta}(E)$ are the charge current 
scattering cross sections of $\nu_\alpha$ and $\nu_\beta$.  Then the total 
number of non-oscillating interacting beam neutrinos at the distance $L$ 
over an energy range is given by
\begin{equation}
N_{0\alpha}(L) = \int N_{0\alpha}(E,L) dE,
\end{equation}
The limits of the energy integration depend on the energy range selected
which can be over the whole energy range of the available beam or a 
selected bin size.

The number of charged leptons of flavor $\beta$ appearing in the detector, 
which is the signal for identifying the appearance of $\nu_\beta$ in the 
$\nu_\alpha$ beam, is given by
\begin{equation}
N_{S\beta}(L) = \int P_{\alpha\rightarrow\beta}(E,L)N_{0\beta}(E,L) dE
\end{equation}
where $P_{\alpha\rightarrow\beta}(E,L)$ is the oscillation probability for
$\nu_\alpha\rightarrow \nu_\beta$.  Again the integration can be over the
whole available energy range of the neutrino beam or a bin size. 
The averaged oscillation probability over the given energy range is  
\begin{equation}
<P_{\alpha\rightarrow\beta}(L)> = {N_{S\beta}(L)\over N_{0\beta}(L)}
\label{eq:probability} 
\end{equation}
where
\begin{equation}
N_{0\beta}(L) = \int N_{0\beta}(E,L) dE
\end{equation}

We now define the figure of merit in terms of the physically measurable
quantity and its uncertainties:
\begin{equation}
F_M(L) \equiv  %(\delta [ln(<P_{\alpha\rightarrow\beta}<L>)])^{-1}
       % = 
           {<P_{\alpha\rightarrow\beta}(L)>\over 
            \delta <P_{\alpha\rightarrow\beta}(L)>}
%{N_{S\beta}(L)\over \delta N_{S\beta}(L)}
\end{equation}
where $\delta <P_{\alpha\rightarrow\beta}(L)>$ denotes the uncertainties of
$<P_{\alpha\rightarrow\beta}(L)>$.
%$\delta N_{S\beta}$ is the error of $N_{S\beta}$. 
In the usual definition this is the number of sigma that determines the quality 
of the measurement. There are in general three sources of uncertainties:
% assuming that the error in all neutrino charge current cross sections 
% can be neglected:
(i) the statistical error in the measurement of the charge lepton of flavor 
$\beta$ which is as usual $\sqrt{N_{S\beta}(L)+f_\beta N_{0\beta}}$,
where $f_\beta N_{0\beta}$ is the number of measured beckground
events, expressed as the fraction $f_\beta$ of $N_{0\beta}$;
%(ii) the statistical error in the measured background events which is
%proportional to $\sqrt{N_{0\alpha}}$ which can be scaled to 
%$\sqrt{N_{0\beta}}$ and will be denoted as $\sqrt{f_\beta N_{0\beta}}$;
%(iii) 
(ii) the systematic uncertainty in the calculation of the number of 
background events from $N_{0\beta}$, which can be denoted as 
$r_\beta f_\beta N_{0\beta}$.  
%(iv)
(iii) the systematic uncertainty in the beam flux and the cross section 
%which will contribute to an error in the signal events 
which we denote as 
$g_\beta N_{S\beta}$.  The total error is the quadrature of all these 
uncertainties. Then the figure of merit can be written as
\begin{eqnarray}
F_M(L) &=& {N_{S\beta}\over
     \sqrt{N_{S\beta} + f_\beta N_{0\beta} + (g_\beta N_{S\beta})^2
           + (r_\beta f_\beta N_{0\beta})^2}}
\nonumber \\
\nonumber \\
  &=& {<P_{\alpha\rightarrow\beta}(L)>\over
 \sqrt{{<P_{\alpha\rightarrow\beta}(L)> + f_\beta \over N_{0\beta}(L)}
           + (g_\beta <P_{\alpha\rightarrow\beta}(L)>)^2 +
             (r_\beta f_\beta)^2 }}.
\label{eq:fom}
\end{eqnarray}
This expression can be generalized to the case when the physical quantity 
under consideration involves two or more averaged probabilities. 
In that situation, each probability has its associated uncertainties.
In the uncertainty terms we will refer to the linear terms which are inversely    %%%
proportional to the beam intensity as the statistical terms, and to the         %%%
quadratic terms as the uncertainty terms.                                                        %%%

The value of $r_\beta$ depends on how well the background is calculated. 
The value of $g_\beta$ depends on how well the beam flux and detection
cross sections are known.
Although $f_\beta$ is not a physical quantity, it can be estimated from the
background of the beam survival measurement.  
%$\nu_{B\beta}$ obtained from the quantity $\nu_\alpha$.  Let us define
%$N_{B\beta} = f^{(0)}_\alpha N_{0\alpha}$.  Then we have
%\begin{equation}
%f_\beta = {N_{0\alpha}\over N_{0\beta}}f^{(0)}_{0\alpha}
%end{equation}
Some of the error factors for different situations have been estimated in
the literature~\cite{errorfactors,bgrw,burguet,freund,yasuda}. Many of
the early treatments included 
only statistical errors; we believe that we have included all 
general uncertainty factors as enumerated above Eq. (\ref{eq:fom}).
In Sec.~5 we will compare our results with previous analyses.
 %In our case we will take
%$r_\beta=0.1$, $g_\beta=0.05$, and $f_{\beta}$=0.01 - 0.03 
%depending on the measurement.  The actual value of the error factor 
%$f_\beta$ will be specified later in each case considered. 

The expression of the figure of merit can be simplified if we restrict the
energy integration to very narrow bins so that the oscillation probability
and the charge current (CC) neutrino cross sections do not vary 
significantly within the bin, then 
$P_{\alpha\rightarrow\beta}(E,L)$ and $N_{0\beta}(E,L)$ are 
essentially constant over the range of integration of the neutrino energy.  
Hence in this narrow bin (NB) approximation we have 
\begin{eqnarray}
F_M(L)_j%|_{\rm NB} 
  &=& {P_{\alpha\rightarrow\beta}(E_j,L)\over
  \sqrt{{{P_{\alpha\rightarrow\beta}(E_j,L) + f_\beta}\over 
               N_{0\beta}(E_j,L)\Delta E}
             + (g_\beta P_{\alpha\rightarrow\beta}(E_j,L))^2
             + (r_\beta f_\beta)^2}}
\label{eq:fombin}
\end{eqnarray}
where $E_j$ is the central value of the energy of the bin and $\Delta E$ 
the bin size, both in GeV.   Generally speaking at shorter
baselines when the oscillation probability is small and the number of
events large, the effect of the background is severe. Typical values
of $f_\beta$ for a water \v{C}erenkov calorimeter are a few
per cent~\cite{H2B}.  For the value of $f_\beta$ of water \v{C}erenkov
ring imaging detector, see Ref.~\cite{J2K}.

We can define the figure of merit over an energy range for a given baseline
length $L$.  It gives the overall statistical significance of the measurement 
over the energy range.  Let the energy range contains N$_b$ bins.  Then, in the 
narrow bin approximation, the total figure of merit in the energy range is 
\begin{equation}
F^{(T)}_M(L) = \sqrt{\sum^{N_b}_{j=1} (F_M(L)_j)^2}
\label{eq:tfombin}
\end{equation}
     
The above expressions can be extended to the case of anti-neutrinos, 
and that containing a mixture of neutrino and anti-neutrino.  The 
quantities associated with the anti-neutrino are defined by replacing the 
neutrino CC cross section by the corresponding anti-neutrino CC
cross section.

\section{Optimum long baselines}                                                     %%%

In this section we apply the Figure of Merit to examine the effectiveness 
of measurements of various relevant quantities at medium and very long
baseline oscillation experiments relevant to the next generation LBL 
experiments. We assume identical detectors at all sites with a neutrino 
beam reaching to the different sites.  
%We consider both types of neutrino beams discussed in the preceding 
%section.  
To limit the scope of the analysis, we will only discuss the 
scenario of three neutrinos with MSW-LMA which is the most favorable 
solution to the solar neutrino problem~\cite{Gonzalez-Garcia}.   
The mixing parameters are set as follows:  % \cite{Valle}:
$\Delta m^2_{32} \approx \Delta m^2_{31}=3.3\times 10^{-3} eV^2$,
$\Delta m^2_{21}=5\times 10^{-5} eV^2$, 
tan$^2\theta_{23}=1.6$, ${\rm tan}^2\theta_{13}=0.025$, and $\theta_{12}=45^0$. 
The leptonic CP phase $\phi$ is set to be 
$90^\circ$.  The earth density has been chosen to be constant of 
$\mathrm \rho=3~g/cm^3$  with equal number of protons and neutrons.
Then the  matter effect constant $A/E_\nu$ is 
$2.3\times 10^{-4} eV^2/GeV$.  

As already stated earlier, for the neutrino beam 
we will take $E_\mu = 20$ GeV for the neutrino factory and $E_p = 50$ GeV 
for the meson beam.  For the error parameters, we will use $r_\beta = 0.1$ 
and $g_\beta=0.05$, independent of the neutrino flavor and beam type.  The 
value of  $f_\beta$, taking the value 0.01 and 0.03, may be different for 
different measurements and neutrino 
beams and will be given explicitly in the various cases considered below.
The beam energy considered ranges from 0.5 to 20 GeV and the 
bin size is $\Delta E = 0.5$ GeV.  
For very long baselines, the cases of neutrino energy of hundreds of MeV  %%%
and lower require separate investigations.  This is because  in such            %%%
lower energy regime, the solar neutrino energy scale will come into            %%%
play and the Eqs. (7) and (8) are on longer valid as the quasi-elastic           %%%
charge interactions \cite{quasi-elastic} become significant.                                                       %%%
The detector size will be taken to
be 100 kt~\cite{100kt} and the running time 3 years.  
For simplicity we use the narrow bin approximation, Eqs.~(\ref{eq:fombin})
and (\ref{eq:tfombin}).  % to calculate the figures of merits. 
We will comment in the Summary and Comments section on neutrino
factories of different primary muon energies.

Before we discuss in detail the individual measurements, we summarize   %%%
the general feature of the contributions of the statistical and uncertainty          %%%
terms.  For the neutrino factory the statistical terms dominate in the           %%%
lower energy region whose range increases with the baseline length.    %%%
Away from the dominant region, the statistical and systematic 
uncertainty terms become comparable with the latter being slightly larger.                                   %%%
For the meson neutrino beam, the statistical terms are dominant in            %%%
regions at both the lower and upper ends of the beam energies.  The         %%%
two types of terms become comparable in the region in between which      %%%
is sizable.  This feature holds even when the beam intensity increases       %%%
or decreases by an order of magnitude from the sample meson                 %%%
neutrino beam intensity considered below.  Hence the statistical                %%%
terms are critical in the determination of the figure of merit.   We can         %%%
conclude that although the individual figures of merit depend on               %%%
the details of the beam intensity and the various error factors, the             %%%
relative figures of merit of different baseline lengths are not too             %%%
sensitive to the details of the beam profile as long as the correct               %%%
neutrino beam energy and baseline distance dependence are               %%%
taken into consideration.  Hence the relative figures of merit discussed     %%%
below should be approximately valid even when the beam intensity                    %%%
profile is not precisely known.                                                                                     %%%

For the total figure of merit as a function of baseline, we
only consider the neutrino factory.  The use of Eq. (6) for the
meson neutrino beam is not meaningful for the total figure of 
merit in view of the complicated nature of the meson neutrino beam.
For comparison we also consider the total figure of merit for 
neutrino factories of 30 and 50 GeV.

To simplify the numerical calculations, we consider only the case of
constant matter density.  We use the vacuum oscillation formulae 
for $\nu_\mu\rightarrow \nu_\mu$, $\nu_\tau$.  For
$\nu_\mu\rightarrow \nu_e$ appearance we use the approximate 
expressions given
in Ref. \cite{approxformu}.  For the CP phase sensitivity we use the 
exact formula given in Ref. \cite{exactformu}.

\subsection{Mixing probability  $\sin^22\theta_{13}$}

The oscillation probability $P(\nu_{\mu}\to\nu_e)$ is a direct measurement 
of $\sin^22\theta_{13}$ since 
$P(\nu_{\mu}\to\nu_e)\propto \sin^22\theta_{13}$.  
%Of course $\sin^22\theta_{13}$ also appears in the $\nu_e\to\nu_\tau$ 
%channel, but experimentally it is much more difficult to apply.  
The statistical 
significance of  $P(\nu_{\mu}\to\nu_e)$ for a given experiment can be 
measured by the {\bf figure of merit}. From Eqs.~(\ref{eq:fom}) and
(\ref{eq:fombin}), we have
\begin{eqnarray}
F^{(1)}_M(L) 
   &=& {<P_{\nu_\mu \rightarrow\nu_e}(L)>\over
       \sqrt{ {<P_{\nu_\mu\rightarrow\nu_e}(L)> +
          f_{\nu_e}\over N_{0\nu_e}(L)}
           + (g_{\nu_\mu} <P_{\nu_\mu\rightarrow\nu_e}(L)>)^2 + 
           (r_{\nu_e} f_{\nu_e})^2 }}
\label{eq:fom1}
\end{eqnarray}

For the error factor $f_{\nu_e}$, contributions from the beam are at the 
level of 0.6\% for meson-neutrino beams.\footnote{See Ref.~\cite{J2K} 
for the water \C ring imaging detector.}  
We conservatively choose $f_{\nu_e}=0.03$ for the conventional meson
beam and $f_{\nu_e}=0.02$ for the neutrino factory. Now the figure of 
merit can be readily calculated using Eqs.~(\ref{eq:fom1}) 
%and (\ref{eq:fombin1}) 
as functions of the neutrino energy E at four 
different baselines: 300 km, 700 km, 2100 km and 3000 km.

To see the importance of the background let us first consider the figure
of merit by ignoring the background, i.e., setting $r=0$, $g=0$ and $f=0$.  
The results are given in Figs.~2(a) and (b).
%Fig.~\ref{fig:s13}a and b. 
As shown in Fig.~2(a),
%Fig.~\ref{fig:s13}a, 
%for a given L,
there is an optimum neutrino energy, $E_{\rm opt}=15$ GeV, independent of
the baseline, where the figure of merit is at a maximum for the neutrino factory.
% and $E_\nu \leq 20$ GeV, there is no 
% $E_{\rm opt}$ where the figure of merit is a maximum for the 
% neutrino factory.  One would be tempted to conclude that it is is more 
% advantageous to work at higher neutrino energies and shorter baselines.  
For the superbeam, each baseline is associated with a different
$E_{\rm opt}$ and the smaller baseline has higher figure of merit 
as shown in Fig.~2(b).
%Fig.~\ref{fig:s13}b.  
In these plots and for all those figure
of merit plots that follow, the dotted line is for $L=3000$ km, the heavy 
dotted line for $L=2100$ km, the dash line for $L=700$ km, and the 
dash-dotted line for $L=300$ km.  For the very long baseline of 
2100 and 3000 km the figures of merit are small and tend to oscillate
rapidly at low energies, so we do not show them for $E < 1.5$ GeV.

We have examined the cases of 30 and 50 GeV neutrino factories and
found that their figures of merit as a function of the neutrino energy are 
similar to those of the 20 GeV case.  The $E_{\rm opt}$ occurs at roughly
the same value and the height of the maximum decreases as the primary 
muon energy increases. 

\begin{figure}
\begin{center}

\vspace{-0.5cm}
\hspace{-1.0cm}
%\mbox{\epsfig{file=s13senb.eps,height=25cm,clip=}}
\mbox{\epsfig{file=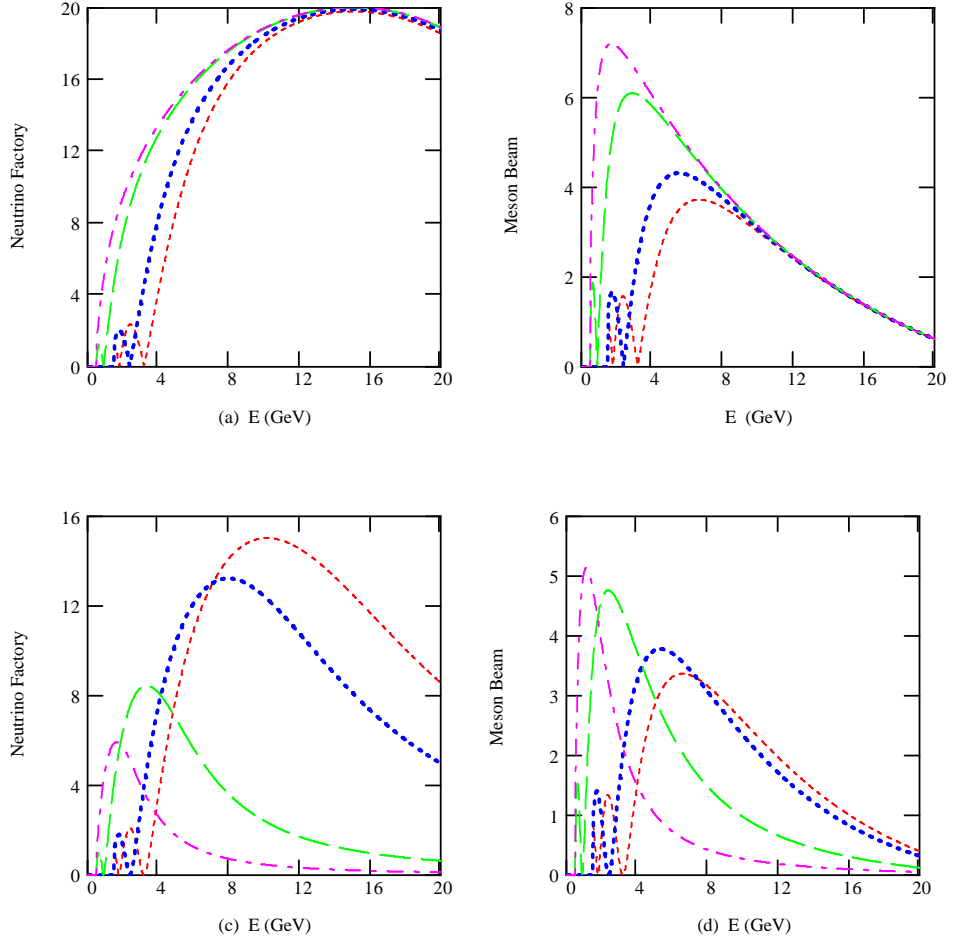,height=25cm,clip=}}
\vspace{-12.0cm}
%\parbox{5.75in}
{\caption{Figure of merit for the $\sin^22\theta_{13}$  
         measurement at a) neutrino factories without backgrounds, b) 
         meson-neutrino beams without backgrounds, c) neutrino factories 
         with f=0.02 and r=0.1, and d) meson-neutrino beams with f=0.03 
         and r=0.1. In these figures and those that follow the dotted line is
         for L=3000 km, the heavy dotted line for L=2100 km, the dash line 
         for L=700 km, and the dash-dotted line for L=300 km.  The 
         2000 and 3000 km curves are not shown below 1.5 GeV. 
        The vertical axis is the figure of merit given by Eq. (\ref{eq:fom1}) 
        and the horizontal axis is the neutrino energy.  }}

\label{fig:s13}
\end{center}
\end{figure}

However, the above picture is changed when the effect of experimental
uncertainties are taken into account. As shown in Figs.~2(c) and (d),
%Figs.~\ref{fig:s13}c and \ref{fig:s13}d, 
the figure of merit is generally reduced. 
%an $E_{opt}$ exists for each distance.   The figure of merit 
For the neutrino factory the figure of merit increases with the baseline 
length, and for the longer distances like 2100 km is much higher than 
the shorter distance like 300 km.  For the conventional neutrino beams
the reduction of the figure of merit at longer distances is less 
when uncertainties are included, and all distances have comparable
figures of merit over the whole energy range.  
Hence the comparison of Figs.~2(c) and (d) with Figs.~2(a) and (b)
%Figs~\ref{fig:s13}c and \ref{fig:s13}d with Figs~\ref{fig:s13}a and 
%\ref{fig:s13}b 
demonstrates the importance of the effect of the background in the 
determination of the relative merits of different baselines.  
It should be noted that the above results are very sensitive to the value 
of ${\rm sin}^2{2\theta_{13}}$.  For a smaller sin$^2{2\theta_{13}}$, the 
effect of the background is larger and the longer baselines will be 
better.
\begin{figure}
%\begin{center}
\centering

\vspace{-1.2cm}
\hspace{-7.0cm}

\mbox{\epsfig{file=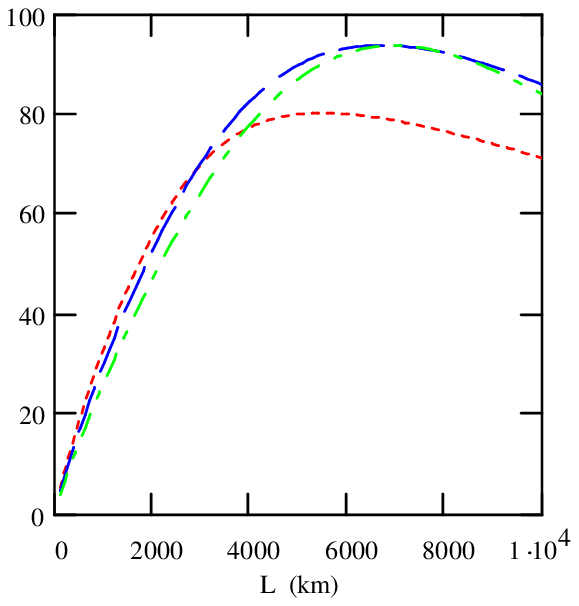,height=25cm,clip=}}
\vspace{-18.5cm}
%\parbox{5.75in}
{\caption{The total figure of merit for $\sin^22\theta_{13}$ as a function
                of baseline length for neutrino factories of 20 GeV (dot), 
               30 GeV (dash), and 50 GeV (dashdot). The horizontal axis is
               the baseline length and the vertical axis is the total figure
               of merit.}}
%      in the energy range of 1-21 GeV.  The dotted curve is for neutrino 
%      factory and the dash curve for the meson beam.}}
\label{fig:As}
%\end{center}
\end{figure}

Figure (3)
%Figure~(\ref{fig:As}) 
gives the total figures of merit, according to Eq. (19), %(\ref{eq: tfombin}), 
of three neutrino factories as functions of 
the baseline: 20, 30 and 50 GeV.  We extend the baseline to
10$^4$ km to show the variation the total figure of merit. 
%t as a function of the dist
%and a meson beam as a function of the baseline length over the energy 
%range 1-21 GeV.  
%The dotted line is for the neutrino factory and the dash
%line for the meson beam.  In order to see the general trend of the variation 
%with the baseline length we extend the plot to 10$^4$ km.  We see that 
%in the case of meson beam the total figure of merit is quickly saturated 
%once the baseline length reaches 1000 km and then decreases very 
%slowly when distance is further increased.  For the neutrino factory the total 
%figure of merit increases with baseline length and becomes saturated at 
%much longer distance, in the present case beyond 6000 km, and then decreases 
%slowly.   In general, the total figure of merit is much smaller at a few hundred 
%km than at longer baseline lengths such as a few thousand km.  The 
%saturation distance are different for different measurements as will be seen 
%in the subsequent discussions of other measurements.   Since we have used 
%a constant matter density, our result will subject to some modification when 
%more realistic, distance varying matter density is used. 
At shorter distance the total figures of merit are roughly the same for all three
neutrino factories.  Their values are small but increase as the baseline 
increases.  The factory of lower energy will reach a maximum sooner
as the baseline increases.   
%The values of the total figure of merit are
%generally significant at and beyond 2000 km, more than 10 except for
%the CP phase.  As will be seen later, for the CP phase of 90$^\circ$ the 
%figure of merit is more than 3 for a 20 GeV factory with the baselines
%between 2000 and 6000 km.   

\subsection{ CP phase $\phi$}

For three flavors there is one measurable phase in the neutrino mixing matrix
which will give rise to a CP-violation effect.  Although the effect of CP phase 
in the hadronic sector is small due to the small Jarlskog factor arising from
small quark mixing angles, the large or even maximal mixing angles of the 
neutrinos suggest that it is not impossible to have a large effect of leptonic 
CP phase. So we will assume the maximal CP phase, $\phi=\pi/2$.
  
The CP phase $\phi$ can be measured by looking at the difference of the 
oscillation probability between $P(\nu_\mu\rightarrow \nu_{e})$ and 
$P(\bar\nu_\mu\rightarrow \bar\nu_{e})$ or between
$P(\nu_e\rightarrow \nu_{\mu})$ and $P(\nu_\mu\rightarrow \nu_e)$.
While the former has to have the matter effect removed in order to obtain
the CP effect, the later, under the assumption of CPT conservation, allows
a direct isolation of the matter from the CP effect, but it can only be
done at a neutrino factory in the case of symmetric matter density
which is approximately true and identical $\nu_e$ and $\nu_\mu$
beam which is generally difficult to arrange.  

We define 
\begin{equation}
\Delta P(E,L,\phi) = P_{\nu_\mu\rightarrow \nu_{e}}(E,L,\phi) -  
                    P_{\bar\nu_\mu\rightarrow \bar\nu_{e}}(E,L,\phi)
\end{equation}
The CP phase can be obtained by subtracting the matter effect
$\Delta P(E,L,\phi) - \Delta P(E,L,0)$. 
Then the figure of merit for the leptonic CP phase measurement can be 
written as 
\begin{eqnarray}
F^{(2)}_M(\phi)
        %&=& {\Delta P(\phi) - \Delta P(0) \over 
        %    \delta(\Delta P(\phi))}   \nonumber \\  
    &=& {<\Delta P(\phi)> - <\Delta P(0)>\over N^{(2)}(\phi)}
   \nonumber  \\
    \nonumber  \\
N^{(2)}(\phi) &=& 
        \left( {<P_(\phi)>+f_{\nu_e}\over N_{0\nu_e}(\phi)} 
               + (r_{\nu_e} f_{\nu_e})^2
          + (r_{\bar{\nu}_e} f_{\bar{\nu}_e})^2 \right.   \\
     & &  \left. + {<\bar{P}(\phi)>+f_{\bar{\nu}_e}\over
                   N_{0\bar{\nu}_e}(\phi)}
          + g^2( <P(\phi)>^2 + <\bar{P}(\phi)>^2) \right) ^{1/2},
          \nonumber
\label{eq:fom2}
\end{eqnarray}
where to simplify the notation we have defined all the quantities by retaining the 
reference to the CP phase and dropping the other variables.

\begin{figure}[htbp!]
\begin{center}

\vspace{0.0cm}
%\vspace{-2.5cm}
\hspace{3.0cm}
%\mbox{\epsfig{file=cpsenb.eps, height=25cm, clip=}}
\mbox{\epsfig{file=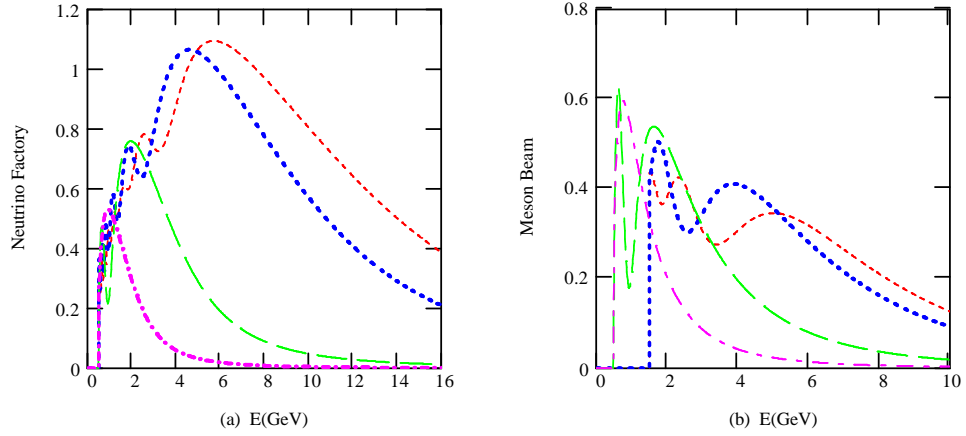, height=25cm, clip=}}
\vspace{-18.cm}
%\parbox{5.75in}
\caption[]{Figure of merit for CP phase measurement at a) 
              neutrino factories with f=0.02 and r=0.1, and 
              b) meson-neutrino beams with f=0.03 and r=0.1. 
              The vertical axis is the figure of merit given by %Eq. (\ref{eq:fom2}) 
              Eq. (22) and the horizontal axis is the neutrino energy.  }
\label{fig:cp12}
\end{center}
\end{figure}

The figures of merit are calculated according to Eq.~(22)
%Eq.~(\ref{eq:fom2}) 
with $f=0.02$ for the neutrino factory and $f=0.03$
for the conventional beam.   Figure~\ref{fig:cp12}(a) shows the figure of 
merit for neutrino factories with the full uncertainties for the neutrino
factory. 
%It is clear from the figure that, for every L, there is an optimum energy 
%for CP phase $\phi$ measurement.
For $L/E$ fixed at the optimum energy, the sensitivity to the CP phase $\phi$ 
is a factor of 2 better at L=2100 km than that at L=300 km for the neutrino
factory.
The sensitivity to the CP phase for the conventional neutrino beam
is shown in Fig.~\ref{fig:cp12}(b). 
A similar optimum energy also exists and the sensitivity at L=2100 km is 
about 20\% lower than that at 
L=300~km at their respective peak values.  
%But the figure merit over the
At neutrino factories, the figure of merit over the whole energy range, i.e.,
the total figure of merit, is 
much larger for 2100 km than for 300 km as shown in Fig.~5.
% for the 20, 30 and 50 GeV factories.

%%%%%%%%%%%%%%%%%%%%%%%%%%%%

\begin{figure}[htbp!]

\begin{center}

\vspace{-1.2cm}
%\vspace{-2.5cm}
\hspace{0.0cm}
%\mbox{\epsfig{file=cpsenb.eps, height=25cm, clip=}}
\mbox{\epsfig{file=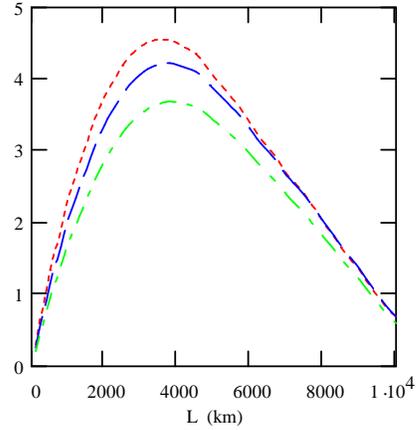, height=25cm, clip=}}
\vspace{-18.cm}
%\parbox{5.75in}
\caption[]{Total figure of merit for CP phase measurement for the neutrino
            factories of 20 GeV (dot), 30 GeV (dash), and 50 GeV (dashdot).
            The horizontal axis is the baseline length and the vertical axis is 
            the total figure of merit. }
%    The dotted curve is for the neutrino factory and dash curve for the 
%    meson beam. }  

\label{fig:Bs}
\end{center}
\end{figure}

%Figure~\ref{fig:Bs} gives the total figure of merit over the energy range of 1-21
%GeV.  It is interesting to note that a broad maximum occurs around 2100 km
%in the meson factory.  For the neutrino factory the total figure of merit peaks 
%slightly below 4000 km and it is relatively sharp.  The figure of merit at 
%2100 km is about 2/3 of the maximum value, while it is about 1/10 at a few
%hundred km.  The values of the total figures of merit and those of the 
%individual energy bins are generally small as demonstrate in 
%Figs.~(\ref{fig:cp12}) and (\ref{fig:Bs}).  Hence for the measurement of this 
%very important quantity it is necessary to increase the statistics by further 
%increasing the neutrino beam luminosity, the detector size and the running 
%time.  We should note again that our result is approximate as we have used 
%a constant matter density. 

\subsection{Sign of $\Delta m^2_{32}$}

Since the vacuum oscillation probability is an even function of 
$\Delta{m}^2_{kj}$, the presence of the matter effect is necessary in 
order to measure the sign of $\Delta m^2_{32}$.  The simplest way is to 
compare the measured probability $P_{\nu_\mu\to\nu_e}$ with the expected 
values of $P^+_{\nu_\mu\to\nu_e}$ and 
$P^-_{\nu_\mu\to\nu_e}$, where $P^+$ assumes $\Delta m^2_{32}>0$ and 
$P^-$ for $\Delta m^2_{32}<0$.  The channel $\bar \nu_\mu\to\bar\nu_e$ 
may be more advantageous to use if $\Delta m^2_{32} <0$.  This
measurement can only be done if $\sin^22\theta_{13}$ is sizable and 
the $\nu_e$ or $\bar\nu_e$ appearance signal is statistically 
significant.  Other channels, such as the $\nu_\mu\to \nu_\mu$ survival 
channel, are less sensitive due to the very small matter dependence.  
It has been suggested in Ref.~\cite{freund2} that the muons from all 
channels at the neutrino factory, e.g. the 
$\nu_e(\bar\nu_e)\to\nu_\mu(\bar\nu_\mu)$ appearance and 
$\nu_\mu(\bar\nu_\mu)\to\nu_\mu(\bar\nu_\mu)$ survival channels, 
be included in a given analysis to reduce systematic errors.  
Here we confine ourselves only to a relatively simple approach
to compare the relative statistical importance at different baselines.
The figure of merit in the present case can be written as
\begin{eqnarray}
F^{(3)}_M(L) 
   &=& {<P^+_{\nu_\mu \rightarrow\nu_e}(L)> - 
        <P^-_{\nu_\mu \rightarrow\nu_e}(L)> \over
       \sqrt{ {<P^+_{\nu_\mu\rightarrow\nu_e}(L)> +
          f_{\nu_e}\over N_{0\nu_e}(L)}
           + (g_{\nu_\mu} <P^+_{\nu_\mu\rightarrow\nu_e}(L)>)^2 + 
           (r_{\nu_e} f_{\nu_e})^2 }}
\label{eq:fom3}
\end{eqnarray}
%and
%\begin{eqnarray}
%F^{(3)}_M(L)|_{\rm NB} 
%  &=& {P^+_{\nu_\mu\rightarrow\nu_e}(E,L)
%       - P^-_{\nu_\mu\rightarrow\nu_e}(E,L) \over
%  \sqrt{{{P^+_{\nu_mu\rightarrow\nu_e}(E,L) + f_e}\over 
%                       N_{0\nu_e}(E,L)\Delta E}
%             + (g_{\nu_\mu} P_{\nu_\mu\rightarrow\nu_e}(E,L))^2
%             + (r_{\nu_e} f_{\nu_e})^2}}
%\label{eq:fombin3}
%\end{eqnarray}

\begin{figure}[htbp]
%\includegraphics[scale=1,width=20cm, height=20cm,
%                          bb=-100 -100 -200 -200]{signb.eps}
\begin{center}

\vspace{-0.5cm}
\hspace{0.0cm}
\mbox{\epsfig{file=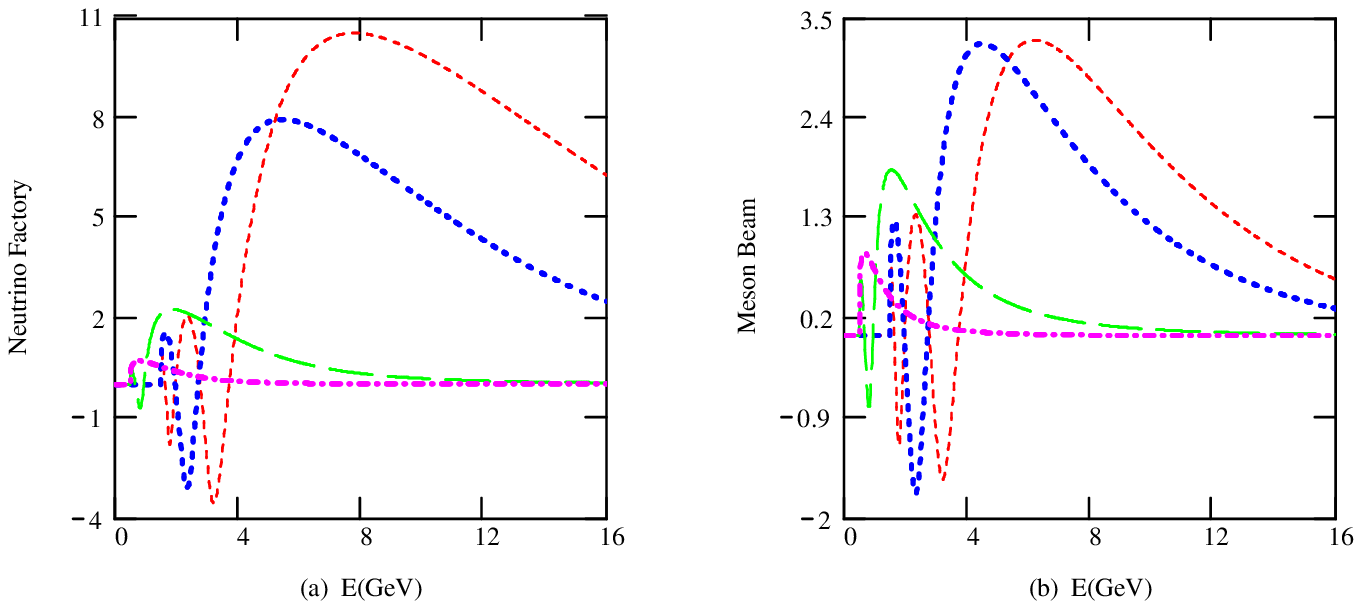,height=25cm,clip=}}
%\mbox{\epsfig{file=signb.eps,height=25cm,clip=}}
%\mbox{\epsfig{file=signb.eps,height=8cm,clip=}}
\vspace{-18.5cm}
%\parbox{5.75in}
{\caption[]{Figure of merit for the sign of $\Delta m^2_{32}$ 
            at a) neutrino factories with f=0.02 and r=0.1, and b) meson-neutrino 
            beams with f=0.03 and r=0.1.
             The vertical axis is the figure of merit given by Eq. (\ref{eq:fom3}) 
        and the horizontal axis is the neutrino energy.  }}
\label{fig:signb}
\end{center}
\end{figure}

%Figure~\ref{fig:signb} 
Figure 6 shows the figure of merit as a function of energy 
at different baselines for neutrino factory and conventional neutrino 
beams with $f=0.02$ for the neutrino factory and $f=0.03$
for the conventional neutrino beam.  It is clear from the plots that for
both beam types, it is much better to have an experiment at longer distance 
such than at shorter distances.  The total figure of merit as a function of 
the baseline length is give in Fig.~7.
%Fig.~\ref{fig:Cs}. 

\begin{figure}[htbp]
\begin{center}

\vspace{-1.5cm}
\hspace{-2.6cm}
\mbox{\epsfig{file=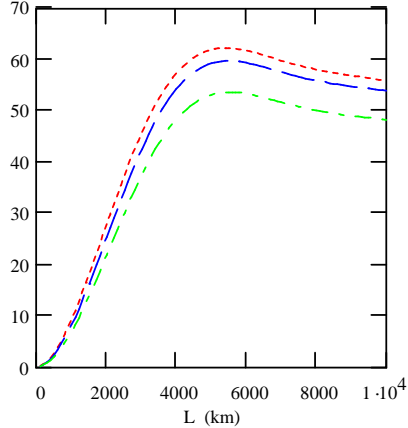,height=25cm,clip=}}
%\mbox{\epsfig{file=signb.eps,height=25cm,clip=}}
%\mbox{\epsfig{file=signb.eps,height=8cm,clip=}}
\vspace{-18.0cm}
%\parbox{5.75in}
{\caption[]{The total figure of merit for the sign of $\Delta m^2_{32}$ 
            for three neutrino factories of 20 GeV (dot), 30 GeV (dash),
            and 50 GeV (dashdot). The horizontal axis is the baseline 
            length and the vertical axis is the total figure of merit.}}
%            over the energy range 1-20 GeV.  The dotted curved is for
%            the neutrino factory and the dash curved for meson beam. }}
\label{fig:Cs}
\end{center}
\end{figure}

\subsection{Matter effect}

The matter effect can be measured by looking at the difference between 
$\nu_\mu\to\nu_e$ and $\bar\nu_\mu\to\bar\nu_e$, although a non-vanishing 
CP phase, $\phi \neq 0$, can alter the result by as much as 20\%. The 
survival channels $\nu_\mu\to\nu_\mu$ and $\bar\nu_\mu\to\bar\nu_\mu$ are 
not very effective for the present purpose since their matter effects 
appear only in nonleading terms. 
%It is better to use the channels that involve $\nu_e$ or $\bar\nu_e$. 

The relevant figure of merit is given by
\begin{eqnarray}
%\begin{eqnarray*}
F^{(4)}_M(L) 
    &=& {<P_1> - <P_2> \over 
        \sqrt{{<P_1>+f \over N_{0\nu_e}(L)} + 2 r^2f^2+
              {<P_2>+f \over N_{0\bar{\nu}_e}(L)} + g^2(<P_1>^2 + <P_2>^2)}},
\label{eq:fom4}
\end{eqnarray}
where we have dropped the explicit reference to all variables except where it 
may cause confusion, and denote 
%$P_1\equiv{P_{\nu_\mu\to\nu_e}(E,L)}$,
%$P_2\equiv{P_{\bar\nu_\mu\to\bar\nu_e}(E,L)}$,
$<P_1>\equiv{<P_{\nu_\mu\to\nu_e}(L)>}$, and
$<P_2>\equiv{<P_{\bar\nu_\mu\to\bar\nu_e}(L)>}$.

%Figure~\ref{fig:matterb} 
Figure 8 shows the figure of merit as a function of energy at different 
baselines for a neutrino factory with $f=0.02$ and the conventional 
neutrino beam with $f=0.03$.  The plots show clearly that  in both cases it is 
much better to have an experiment at longer baseline than at a shorter one.  
As show in 
%Fig.~(\ref{fig:Ds}) 
Fig.~9, the total figures of merit as a function of the 
baseline length for three neutrino factories, 20, 30 and 50 GeV.
% is similar to that in the case of the measurement of the sign 
%of $\Delta{m}^2_{32}$, Fig.~(\ref{fig:Cs}).  However in the present case
%the more realistic Earth matter density will be necessary in a refined calculation.

\begin{figure}[htbp]
\begin{center}

\vspace{-0.5cm}
\hspace{0.0cm}
%\mbox{\epsfig{file=matterb.eps,height=25cm,clip=}}
\mbox{\epsfig{file=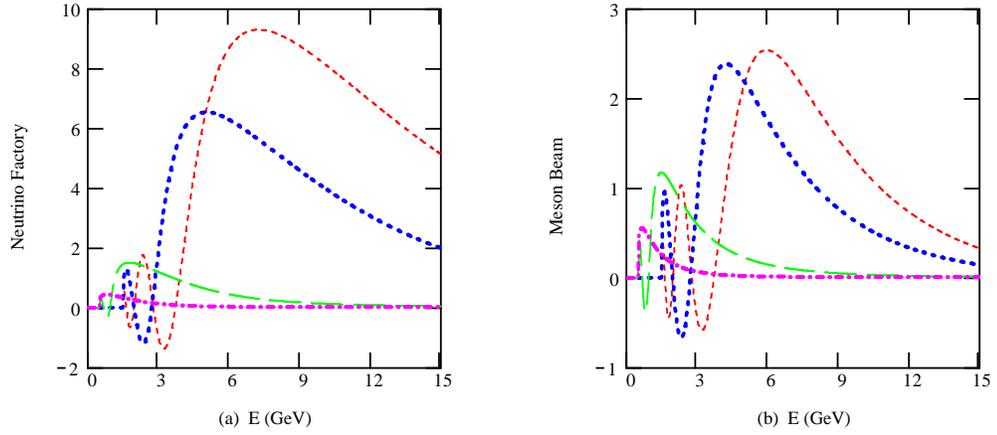,height=25cm,clip=}}
\vspace{-18.5cm}
%\parbox{5.75in}
{\caption[]{Figure of merit for the matter effect at a) neutrino 
                  factories with f=0.02 and r=0.1, and b) meson-neutrino beams 
                  with f=0.03 and r=0.1.
                  The vertical axis is the figure of merit given by Eq. (\ref{eq:fom4}) 
                   and the horizontal axis is the neutrino energy.  }}
\label{fig:matterb}
\end{center}
\end{figure}

\begin{figure}[htbp]
\begin{center}

\vspace{-1.5cm}
\hspace{-2.6cm}
\mbox{\epsfig{file=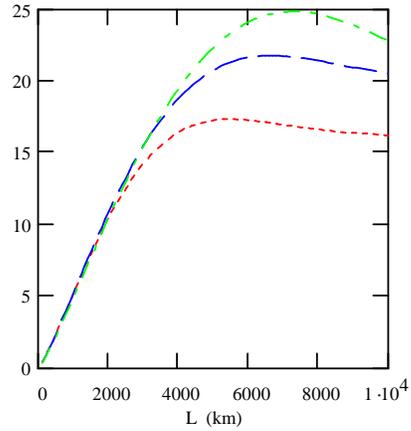,height=25cm,clip=}}
%\mbox{\epsfig{file=signb.eps,height=25cm,clip=}}
%\mbox{\epsfig{file=signb.eps,height=8cm,clip=}}
\vspace{-18.0cm}
%\parbox{5.75in}
{\caption[]{The total figure of merit of the matter effect for three neutrino
                  factories:  20 GeV (dot), 30 GeV (dash), and 50 GeV (dashdot).
                  The horizontal axis is the baseline length and the vertical axis 
                   is the total figure of merit.}}
%         over the energy range 1-20 GeV.  The dotted curved is for
%           the neutrino factory and the dash curved for meson beam. }}
\label{fig:Ds}
\end{center}
\end{figure}

\subsection{Precision measurement of $\Delta m^2_{32}$ and
               $\sin^22\theta_{23}$}

Although $\Delta m^2_{32}$ and $\sin^22\theta_{23}$ have been measured at
Super-K  and hopefully will be further improved by K2K, MINOS, OPERA and 
ICARUS, it is still interesting  to measure them at different 
baselines and possibly improve the precision. 
$\Delta m^2_{32}$ and $\sin^22\theta_{23}$ can be directly related
to the survival probability $P(\nu_\mu\to\nu_\mu)$.
The figure of merit can be written as 
\begin{eqnarray}
F_M^{(5)}(L) = {1-<P>\over\sqrt{ {<P>+f\over N_{0\nu_\mu}(L)} + r^2 f^2 +
                              (g<P>)^2} },
%           \\
%F^{(5)}(L)|_{\rm NB} = {1-P\over\sqrt{ {P +f\over
%                              N_{0\nu_\mu}(E,L)\Delta E} 
%                             + r^2 f^2 + (g P )^2} },
\label{eq:fom5}
\end{eqnarray}
where $<P>\equiv <P_{\nu_\mu\rightarrow\nu_\mu}(L)>$.
% and $P \equiv P_{\nu_\mu\rightarrow\nu_\mu}(E,L)$.

%Figure~\ref{fig:m23b} 
Figure 10 shows the figure of merit at different baselines  for neutrino 
factories and conventional neutrino beams. Since the background for 
muon identification is generally much smaller than in 
the case of the electron, we use $f=0.01$ for both the neutrino factory
and the conventional neutrino beam.  A distinctive peak corresponding to 
the value of 
$\Delta m^2_{32}L$ where the oscillation is at a maximum can be found 
and the position of the peak is a good measurement of $\Delta m^2_{32}$. 
Here the figure of merit can be large for all energies with longer distances
favoring measurements at higher energies.  
%It is clear that for both the neutrino factories and the meson beam it 
%is better at 2100 km than at 300 km.  
The total figure of merit as shown in Fig. 11.
%Fig.~(\ref{fig:Es}) has a relatively sharp peak beyond 7000 km in the 
%case of the neutrino factory.  The meson beam favors the baseline 
%lengths in the range of 1000 to 3000 km.

\begin{figure}[htbp]
\begin{center}
\vspace{-.5cm}
\hspace{-2.0cm}

%\mbox{\epsfig{file=m23b.eps,height=25cm,clip=}}
\mbox{\epsfig{file=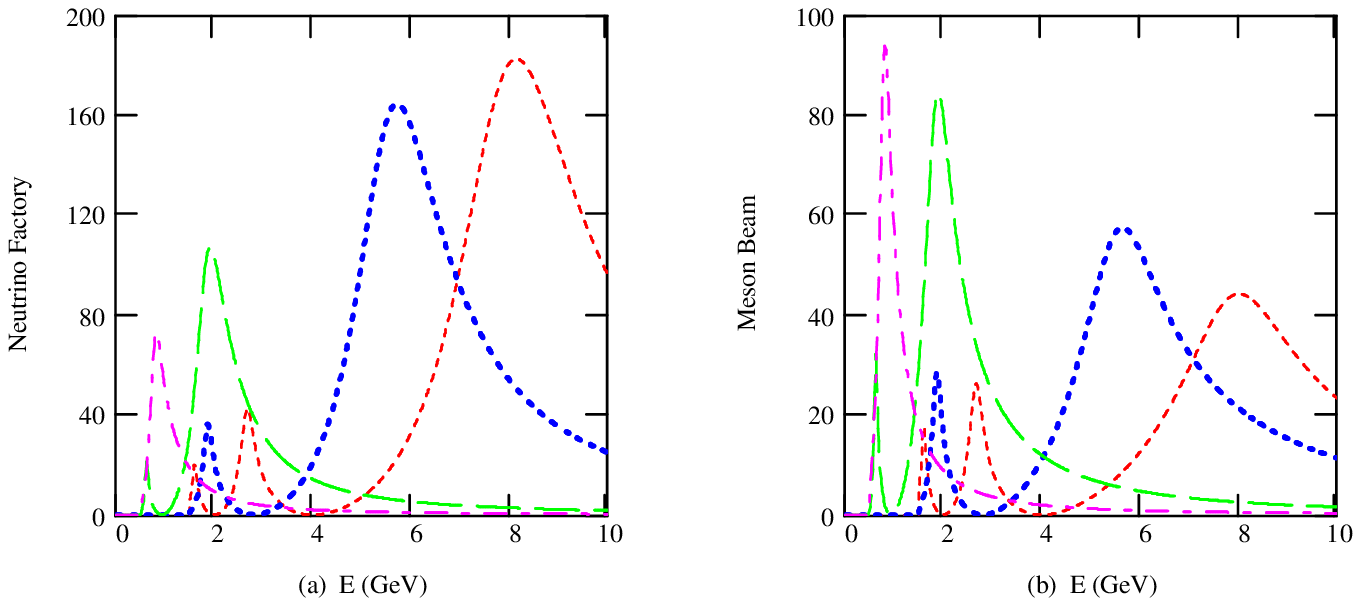,height=25cm,clip=}}

\vspace{-18.5cm}
%\parbox{5.75in}
{\caption[]{Figure of merit for $\Delta m^2_{32}$ and 
                  $\sin^22\theta_{23}$ at a) neutrino factories with f=0.01 and 
                  r=0.1, and b) meson-neutrino beams with f=0.01and r=0.1.
                  The vertical axis is the figure of merit given by Eq. (\ref{eq:fom5}) 
                  and the horizontal axis is the neutrino energy.  }}
\label{fig:m23b}
\end{center}
\end{figure}

\begin{figure}[htbp]
\begin{center}

\vspace{-1.5cm}
\hspace{-2.6cm}
\mbox{\epsfig{file=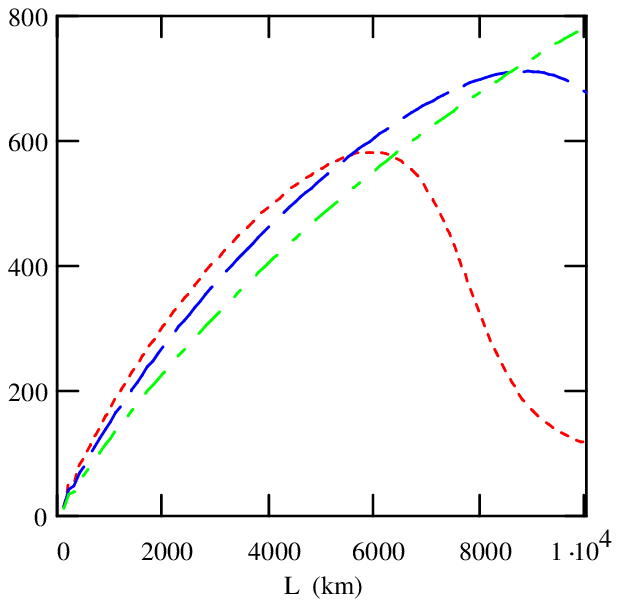,height=25cm,clip=}}
%\mbox{\epsfig{file=signb.eps,height=25cm,clip=}}
\vspace{-18.0cm}
%\parbox{5.75in}
{\caption[]{The total figure of merit for $\Delta m^2_{32}$ and 
            $\sin^22\theta_{23}$ for the three neutrino factories:
            20 GeV (dot), 30 GeV (dash), and 50 GeV (dashdot).
            The horizontal axis is the baseline length and the vertical 
            axis is the total figure of merit.}}
%         over the energy range 1-20 GeV.  The dotted curved is for
%         the neutrino factory and the dash curved for meson beam. }}
\label{fig:Es}
\end{center}
\end{figure}

\subsection{$\nu_\mu\to\nu_\tau$ Appearance}

The appearance of $\nu_\tau$ from a $\nu_\mu$ beam is an unambiguous 
signal of the $\nu_\mu\to\nu_\tau$ oscillation. Precision measurement 
of this oscillation probability is crucial in establishing the 
oscillation patterns and in determining whether or not 
$\nu_\mu\to\nu_\tau$ is the only dominant oscillation mode or if there 
is still room for the $\nu_\mu\to\nu_s$ oscillation, where $\nu_s$ 
is a sterile neutrino.  Although we expect that in the next few years, 
K2K and/or OPERA will observe the production of $\tau$ because of the 
large $\nu_\mu\to \nu_\tau$ probability, it is desirable that future 
detectors possess good $\tau$ identification capability. 
%We assume that our detector can identify the $\tau$ based on 
%statistics methods,  namely event selection via signiture on kinematics. 
The figure of merit can be written as 
\begin{eqnarray}
F_M^{(6)}(L) &=& {<P_{\nu_\mu\rightarrow\nu_\tau}(L)> \over 
      \sqrt{{<P_{\nu_\mu\rightarrow\nu_\tau}(L)> +f \over N_{0\nu_\tau}(L)} 
                    + r^2f^2 + (g<P_{\nu_\mu\rightarrow\nu_\tau}(L)>)^2}}
% \\   
%F^{(6)}(L)|_{\rm NB} &=& {P_{\nu_\mu\rightarrow\nu_\tau(E,L)} \over 
%      \sqrt{{P_{\nu_\mu\rightarrow\nu_\tau(E,L)} +f \over 
%                            N_{0\nu_\tau}(E,L)\Delta E} 
%                    + r^2f^2 + (g P_{\nu_\mu\rightarrow\nu_\tau}(E,L))^2 }}.
\label{eq:fom6}
\end{eqnarray}

%Figure~\ref{fig:mtauc} 
Figure 12 shows this figure of merit as a function of energy 
at different baselines.
% for neutrino factories and conventional neutrino beams. 
It is expected that the error in the present case will be larger than measurements 
of other physical variables due to $\tau$ event selection.  Hence we set f=0.03 
for both neutrino factories and meson-neutrino beams.  Due to the threshold of 
tau production, we made a cutoff at $E_\nu = 4$ GeV.
%the experiment has to be done at $\mathrm E_{\nu_\mu}>4 GeV$.   
For the neutrino factory the larger distance 
is always better.  For the meson beam 2100 km is also better than 300 km for
neutrino energy above 6 GeV.

The total figure of merit is shown in Fig. 13.   %Fig.~(\ref{fig:Fs}).  
As shown, for $L > 1000$ km the total figure of merit favors factory of
higher energy.   However, since the figure of merit of all factories are
high, it is not necessary to maximize the figure of merit.  A 20 GeV
neutrino factory is more than adequate.
%It is interesting to
%note that the maximum in the case of neutrino factory occurs at around 
%2000~km and at 1000~km for the meson beam.  In both neutrino beams
%the 2100~km baseline length is much better than a few hundred km.

\begin{figure}[htbp]
\begin{center}

\vspace{-0.5cm}
\hspace{0.0cm}
%\mbox{\epsfig{file=mtaub.eps,height=25cm,clip=}}
\mbox{\epsfig{file=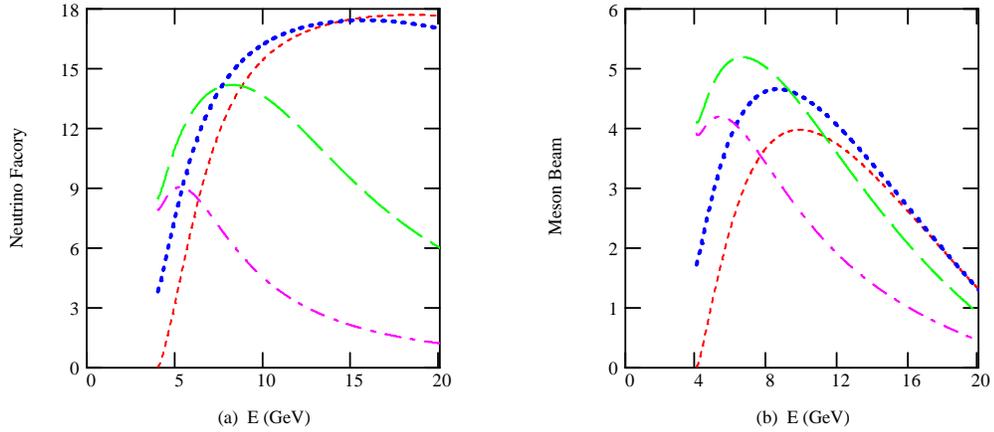,height=25cm,clip=}}
\vspace{-18.5cm}
%\parbox{5.75in}
{\caption[]{Figure of merit for tau appearance at a) neutrino 
                  factories with f=0.03 and r=0.1, and b) meson-neutrino beams 
                   with f=0.03 and r=0.1.
                  The vertical axis is the figure of merit given by Eq. (\ref{eq:fom6}) 
                  and the horizontal axis is the neutrino energy.  }}
\label{fig:mtauc}
\end{center}
\end{figure}

\begin{figure}[htbp]
\begin{center}

\vspace{-1.5cm}
\hspace{-2.6cm}
\mbox{\epsfig{file=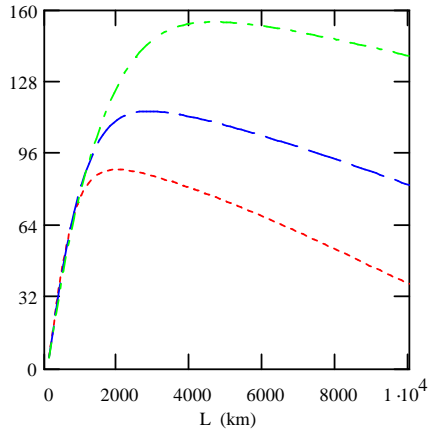,height=25cm,clip=}}
%\mbox{\epsfig{file=signb.eps,height=25cm,clip=}}
%\mbox{\epsfig{file=signb.eps,height=8cm,clip=}}
\vspace{-18.0cm}
%\parbox{5.75in}
{\caption[]{The total figure of merit for the $\tau$ appearance measurement
            for the three neutrino factories: 20 GeV (dot), 30 GeV (dash), and
            50 GeV (dashdot).  The horizontal axis is the baseline length 
            and the vertical axis is the total figure of merit.}}
%         over the energy range 1-20 GeV.  The dotted curved is for the 
%         neutrino factory and the dash curved for meson beam. }}
\label{fig:Fs}
\end{center}
\end{figure}

\section{Summary and Comments}

In Table~2 %\ref{tab:summ} 
we list the maximal values of all the figures of 
merit for all the measurements considered in the range 0.5 to 20 GeV for
a neutrino factory of 20 GeV muon and a superbeam of 50 GeV primary
proton.  To put things in perspective, a few remarks are due.  The value 
of the figure of merit is equal to the same value of $\sigma$ in statistical
significance.  For a meaningful measurement we need to have a figure of
merit no less than 3.  Since the values of the figure of merit are obtained 
in this paper with explicit assumptions about the neutrino flux, the detector 
size,  and the running time, we cannot take the values in the table too literally.  
However, as discussed in the preceding section, the relative merits of  
different baseline, which are not too sensitive to the explicit 
experimental conditions, are meaningful.  The table shows that a longer 
baseline is preferred over a shorter one for the neutrino factory.
For the meson neutrino beams, the preference depends on the variables
measured 
%longer distance is also general preferred, 
but longer baseline length is generally slightly more preferred.  For a 
more complete measurement
of all the relevant variables, the combination of a relatively short and long 
baselines, such as 300 and 2100 km, will do a much better job.  

The CP phase measurement is a challenging case.  The values of the
associated figure of merit at individual energies are generally small in 
all the cases we have considered.   This will be true even for the total
figure of merit in the case of superbeam.  For the neutrino factory the
value of the total figure of merit is already reasonably large at 2100 km
for the 20 GeV factory.  Generally speaking it is desirable to improve
the statistics by improving the running conditions in all aspects:  the 
beam luminosity, the running time, and the detector size.  
 
%while the maxima of the figures of merit at 300 km and 
%2100 km are comparable, the total figure of merit at 2100 km is much higher 
%than that at 300 km.  However, the values of the figures of merit are generally 
%low at all the distances considered under the assumed running conditions.  So 
%improved statistics made by improved running conditions in all the three 
%aspects:  the beam luminosity, the running time and the detector size, are 
%needed for the CP measurement.

\begin{table}[hbtp!]
\begin{center}
\begin{tabular}{|l|c|c|c|c|}
\hline
            & \multicolumn{2}{c|}{neutrino factory}  & 
            \multicolumn{2}{c|}{meson-neutrino beam} \\
            & 300 km & 2100 km & 300 km & 2100 km \\
\hline
$\sin^22\theta_{13}$            & 6.0      & 13      & 5.2      & 3.9     \\
CP phase $\delta$                & 0.55     & 1.1    & 0.6     & 0.5        \\
sign of $\Delta m^2_{32}$   & 0.7      & 8.0     & 1.0     & 3.2    \\
 matter effects                     & 0.5      & 6.6     & 0.6      & 2.4  \\
$\Delta m^2_{32}$ and $\sin^22\theta_{23}$ 
                                           & 75      & 165     & 95       & 58     \\ 
$\tau$ appearance                & 9        & 17.5    & 4.2      & 4.7   \\ 
\hline
\end{tabular}
\parbox{5.75in}{\caption[]{Summary of relative figures of merit (at the
                            maxima) for various measurements at the baselines 
                            L=300 and 2100 km.}}
\label{tab:summ}
\end{center}
\end{table}

%Since the neutrino beam has generally broad energy distribution, the more
%realistic figures of merit should be integrated over the expected energy 
%spectrum with the inclusion of a realistic distance vary Earth matter density.
%While for the neutrino factory, the neutrino energy spectrum is well represented 
%by Eq.(\ref{eq:nuf}), it is more complicated for meson-neutrino beams since 
%specific beam line design can alter significantly the neutrino beam energy 
%spectrum.  As a general rule of thumb it it is always better to have a larger 
%width around the maximum of the figure of merit curve.  To this respect, it 
%is also advantageous at longer baselines is always preferred than, say, 300 km.

Since the flux of the neutrino factory given in Eq. (5) is realistic, it is 
appropriate to make some 
detailed observation in the case of neutrino factory on the dependence of 
the figure of merit on the primary muon energy and baseline length:

(a) First consider the figure of merit at a fixed distance as a function of 
the neutrino energy.  From the 
neutrino flux of the neutrino factory, Eq. (\ref{eq:neutrinofactory}), it can 
be seen that for a given neutrino energy and $n^{(f)}_0$, the neutrino flux 
decreases with increasing $E_\mu$ for $E_\mu > 4E_\nu/3$.  Therefore at 
a given baseline, the maxima of the figures of merit of the various 
measurements moves to higher neutrino energies, in all cases very slowly,
when $E_\mu$ increases, while the height of the maximum generally
decreases significantly.  This is born out in our investigation of the 
30 GeV and 50 GeV factories.  
  
%Therefore, in a fixed range of neutrino energy, the neutrino factory of lower 
%primary muon energy is preferred. \\
(b) Next consider the total figure of merit as a function of the baseline
obtained by integrating the figure of merit over the whole available 
energy range at a fixed baseline.  The total figure of merit 
increases almost linearly with baseline, starting at small baseline, with 
the values of the total figure of merit about the same for all three 
factories.  The values of the figure of merit of the three factories will 
separate with further increases in the baseline.  The lower energy 
factory will reach a maximum first. All the maxima are very broad.   
The first maximum in all the variables is reached by the 20 GeV 
factory at about 2000 km for the tau appearance experiment.  

(c) All the values of the total figure of merit for the neutrino factories
we considered are very significant, larger than 10, for baseline beyond
2000 km, except for the CP phase.  The figure of merit for the
CP phase of 90$^\circ$ has maxima in the 3500 km and 4000 km 
region.   The height of the maximum decreases with increasing 
neutrino factory energy.  The 20 GeV factory has the maximum 
figure of merit value of 4.5 at 3500 km and at 2000 km it is 3.6, 
which are reasonably large.   

(d) Based on the above two considerations the lower energy factory of
20 GeV is preferable considering its lower budget of construction and
possibly higher physics output.  However for a more complete 
investigation, a detailed simulation on the actual number of events
should be performed.

(e) All of the studies on neutrino factories use similar
principles in determining the best $L$ and $E$ combination, i.e., there is
some goodness of fit criterion. The differences are in the details, such as
which sources of uncertainties are included, the particular values
assigned, and whether or not error correlations are taken into account.
Our total figure of merit for neutrino factories gives results
similar to those obtained in other studies. For example, we find that
the optimum $L$ for detecting a nonzero $\sin^22\theta_{13}$ in a
neutrino factory is about 6000--7000~km, and that higher stored muon
energies do better; similar conclusions are reached in
Ref.~\cite{freund,yasuda}. The optimum $L$ for detection of $CP$
violation is 3000--4000~km, which agrees with similar
studies~\cite{errorfactors,bgrw,burguet,freund,yasuda}. We find that the
optimum $L$ for determining the sign of $\Delta m^2_{32}$ is around
5000~km, consistent with the conclusions of Refs.~\cite{bgrw} and
\cite{yasuda}. However, we have also systematically applied the figure 
of merit to other oscillation variables, such as the matter effect, 
$\nu_\tau$ appearance and $\theta_{23}$, most of which have not been
applied before. Furthermore, we have made a
comparison of neutrino factories and meson neutrino beams using identical
criteria.

Although we have not considered the total figure of merit for the case of
the meson neutrino due to the lack of generally valid beam energy profile, 
we can nevertheless draw some useful conclusions based on the salient 
features of the differential figure of merit.  The maximum of the figure 
of merit at a shorter baseline tend to be sharply peaked, while that of a 
longer baseline is broad.  Hence the total figure of merit integrated 
around the maximum will favor longer baselines.  Naively integrating the 
differential figure of merit over the envelope of the beam intensity, Eq. (6), 
shows that the values at 2100 km are larger than at 300 km.  

Finally, we note that in the present work, we focus on the statistical 
significance of long  and very long baselines in the energy range of 
0.5-20 GeV.  
Consequently the dominant mass scale is the atmospheric neutrino scale, i.e., 
$\Delta{m}^2_{32}$.  For neutrino energies of a few hundred MeV or smaller, 
the effect of the solar neutrino scale becomes important. Therefore a
separate investigation is necessary.  

\vskip 4ex

\noindent
{\bf {\large Acknowledgement}}\\

We would like to thank Dr. Xinmin Zhang and Dr. Lianyou Shan for
helpful assistance.   We would also like to thank our colleagues
of the H2B collaboration \cite{H2B} for support.  This work is
supported in part by DOE Grant No. DE-FG02-G4ER40817.

\newpage
%\specialchapt{Bibliography}

\end{document}